\documentclass[aps,preprint,showpacs,floatfix,preprintnumbers,groupedaddress,nofootinbib]{revtex4}
\usepackage{graphicx}
\usepackage{epsfig}
\preprint{\vbox{%
\hbox{\bf YITP-SB-09-005}}}

\begin{document}

\vspace*{.25in}

\title{Identifying Galactic PeVatrons with Neutrinos}
\author{M.C.~Gonzalez-Garcia}
\email{concha@insti.physics.sunysb.edu}
\affiliation{%
  C.N.~Yang Institute for Theoretical Physics,
  SUNY at Stony Brook, Stony Brook, NY 11794-3840, USA
  \\
  Instituci\'o Catalana de Recerca i Estudis Avan\c{c}ats (ICREA),
  Departament d'Estructura i Constituents de la Mat\`eria, Universitat
  de Barcelona, 647 Diagonal, E-08028 Barcelona, Spain}
\author{Francis Halzen}
\email{flhalzen@facstaff.wisc.edu}
\affiliation{%
Department of Physics, University of Wisconsin, Madison, WI 53706,USA}
\author{Soumya Mohapatra}
\email{soumya@insti.physics.sunysb.edu}
\affiliation{%
  C.N.~Yang Institute for Theoretical Physics,
  SUNY at Stony Brook, Stony Brook, NY 11794-3840, USA\vspace*{.25in}}

\begin{abstract}
We perform a realistic evaluation of the potential of IceCube,  
a kilometer-scale neutrino detector under construction at the South Pole, 
to detect neutrinos in the direction of the potential accelerators of
the Galactic cosmic rays. We take fully account of the fact
that the measurement of the energy of the secondary muons can be used
to further discriminate between the signal and the 
background of atmospheric neutrinos. We conclude that IceCube 
could identify the sources
in the Milagro sky map as the sources of the Galactic cosmic rays at
the $3\sigma$ level in one year and at the $5\sigma$ level
in three years. We discuss the dependence of these expectations on ambiguities, mostly associated with our incomplete knowledge of the astrophysics of the sources.
\end{abstract}


\maketitle

\section{Resolving the Galactic Cosmic Ray Problem}

In 1934, Baade and Zwicky\cite{zwicky} pointed out that supernovae
could be the sources of the Galactic cosmic rays provided that a
substantial fraction of the energy released in the explosion is
converted into the acceleration of relativistic particles. Their
proposal has been commonly accepted despite the fact that to date no
source has been conclusively identified. Because cosmic rays are
deflected by magnetic fields, the sources must be identified by the
accompanying pionic gamma rays and neutrinos produced when the
accelerated particles interact with Galactic hydrogen. 

Independently of the nature of the cosmic ray accelerator, it is a
fact that galactic cosmic rays reach energies of at least several PeV,
the ``knee" in the spectrum; their interactions should
generate gamma rays and neutrinos from the decay of secondary pions
reaching hundreds of TeV. Such sources should be identifiable by a hard spectrum that extends to hundreds of TeV without attenuation;
they have been dubbed PeVatrons. Indeed as we will emphasize in the
next section, straightforward energetics arguments are sufficient to
conclude that present air Cherenkov telescopes should have the
sensitivity to detect TeV photons from PeVatrons. In fact, they may
have been revealed by an all-sky survey in the 10\,TeV
energy region with the Milagro detector\cite{Abdo:2006fq}. A subset of
sources located within nearby star forming
regions in Cygnus and in the vicinity of galactic latitude
$l=40$\,degrees are identified that are not readily associated with known supernova remnants or with
non-thermal sources observed at other wavelengths. 
Subsequently directional air Cherenkov telescopes
were pointed at 3 of the sources revealing them as potential PeVatrons
with an approximate $E^{-2}$ energy spectrum that extends to tens of
TeV without evidence for a cutoff\, \cite{hesshotspot,magic2032}.
This is in sharp contrast with the best studied supernova remnants 
RX J1713-3946 and  RX J0852.0-4622 (Vela Junior).

The Milagro sources are suspected to be molecular clouds illuminated
by the cosmic ray beam accelerated in young remnants located within $\sim 100$\,pc.  
In brief, one expects that multi-PeV cosmic rays are accelerated only 
during a short
period when the remnant transitions from the free-expansion to the
beginning of the Sedov phase and the shock velocity is high. The high
energy particles can produce photons and neutrinos over much longer
times as they diffuse through the interstellar medium to interact with
nearby molecular clouds; for a detailed discussion see
reference~\cite{gabici}. An association of molecular clouds and supernova remnants is expected in star forming regions.

Unfortunately the basic hurdle to conclusively associate the observed
TeV gamma rays with the decay of pions produced by a cosmic
accelerator has not been overcome. Synchrotron radiation by energetic
electrons followed by inverse Compton scattering, routinely observed
in non-thermal sources, cannot be excluded as their origin.  Neutrinos
from the decay of charged pions accompanying pionic gamma rays can
provide incontrovertible evidence for cosmic ray acceleration in the
source. It is the purpose of this paper to assess as realistically as
possible the actual prospects for detecting them.

The outline of the paper is at follows. We first set the
stage by selecting the TeV sources that can be reasonably expected to
be candidate Galactic cosmic ray accelerators, i.e. supernova remnants
or associated molecular clouds. We will subsequently derive the
neutrino flux from data collected by TeV gamma ray
detectors. Especially for the case of molecular clouds, the neutrino
flux should be predictable at a quantitative level as any confusion of
pionic gamma rays with gamma rays of electromagnetic origin should be
minimal\cite{gabici}. 
This argues for a more realistic evaluation of
the potential of the IceCube detector taking fully account of the fact
that the measurement of the energy of the secondary muons can be used
to further discriminate between the signal and the background of atmospheric neutrinos.

Our results agree with previous estimates \cite{Halzen:2008zj} that a
neutrino signal should emerge after several years from the data of a
kilometer-scale detector with the predicted fluxes at the level of
the background. For average values of the
parameters we find that the completed IceCube detector could confirm
the sources in the Milagro sky map as sites of cosmic ray acceleration at the $3\sigma$ level in less than one year and at the
$5\sigma$ level in three years. We discuss the considerable dependence
of these expectations on the remaining ambiguities in this estimate. With the performance of IceCube relatively well understood, these are mostly of astrophysical origin. The absence of evidence
after 10 years would definitely reopen the question of the origin of Galactic
cosmic rays and represent an interesting challenge. Whereas other
speculations for their origin exist, for instance micro-quasars, it is
difficult to reconcile them with the fact that they do not appear in
the Milagro sky.

Finally we briefly comment on the prospects for a northern detector
such as KM3NeT viewing southern sources. Opportunities are actually
limited; at this time there are only two candidate sources, RX
J1713-3946 and Vela Junior. It is important to point out however that
the northern sources viewed by IceCube were only revealed after the
commissioning of all sky TeV instruments such as Milagro. No such
instrument has studied the Southern hemisphere and, for the near
future, the only opportunity to do so is to use IceCube as a gamma ray
detector. A quantitative evaluation of this possibility is in
progress\cite{aongus}.

\section{Candidate Galactic Accelerators}

The energy density of the cosmic rays in our Galaxy is $\rho_{E} \sim
10^{-12}$\,erg\,cm$^{-3}$. Galactic cosmic rays are not forever; they
diffuse within the microgauss fields and remain trapped for an average
containment time of $3\times10^{6}$\,years. The power needed to
maintain a steady energy density requires accelerators delivering
$10^{41}$\,erg/s. This happens to be 10\% of the power produced by
supernovae releasing $10 ^{51}\,$erg every 30 years ($10 ^{51}\,$erg
correspond to 1\% of the binding energy of a neutron star after 99\%
is initially lost to neutrinos). This coincidence is the basis for the
idea that shocks produced by supernovae exploding into the
interstellar medium are the accelerators of the Galactic cosmic rays.

Despite the rapid development of instruments with improved
sensitivity, it has been impossible to conclusively pinpoint supernova
remnants as the sources of cosmic rays by identifying accompanying
gamma rays of pion origin. A generic supernova remnant releasing an
energy of $W\sim10^{50}\,$erg into the acceleration of cosmic rays
will inevitably generate TeV gamma rays by interacting with the
hydrogen in the Galactic disk. The emissivity in pionic gamma rays
$Q_{\gamma}$ is simply proportional to the density of cosmic rays
$n_{cr}$ and to the target density $n$ of hydrogen atoms. Here $n_{cr}
\simeq 4\times10^{-14}\, {\rm cm}^{-3}$ is obtained by integrating the 
proton spectrum for energies in excess of 1 TeV. For a $E^{-2}$ spectrum,
\begin{equation}
Q_\gamma  \simeq  c \left<{E_\pi \over E_p}\right> {\lambda_{pp}}^{-1}\, n_{cr}\ ({>}1\,{\rm TeV})
\simeq 2 c x_{\gamma}  \sigma_{pp}\, n\, n_{cr} \; , 
\label{eq:qgam}
\end{equation}
or 
\begin{equation}
Q_\gamma (> 1\,{\rm TeV}) \simeq 10^{-29} \,{\rm photons\over \rm cm^3\,s}\, \left({n \over \rm 1\,cm^{-3}}\right),
\end{equation}
The proportionality factor in Eq.~(\ref{eq:qgam}) is determined by particle 
physics; $x_{\gamma}$ is the average energy of secondary photons relative to
the cosmic ray protons and $\lambda_{pp}= (n\sigma_{pp})^{-1}$ is the
proton interaction length ($\sigma_{pp} \simeq 40$\,mb) in a density
$n$ of hydrogen atoms. The corresponding luminosity is
\begin{equation}
L_{\gamma} ({>} 1\,{\rm TeV})  \simeq Q_{\gamma}\, {W \over \rho_E} 
\end{equation}
where $W/\rho_E$ is the volume occupied by the supernova remnant. 
We here made the approximation that the volume of the young remnant is
approximately given by $W/\rho_E$ or, that the density of particles in
the remnant is not very different from the ambient energy density
$\rho_E \sim 10^{-12}$\,erg\,cm$^{-3}$ of Galactic cosmic rays.

We thus predict a rate of TeV photons from a supernova at a nominal
distance $d$ of order 1\,kpc~of
\begin{equation}
\int_{E>1{\rm TeV}} \frac{dN_\gamma}{dE_\gamma} dE_\gamma= {L_\gamma (>{\rm 1 TeV}) \over 4\pi d^2}
\simeq 10^{-12}-10^{-11} \left({\rm photons\over \rm cm^2\,s}\right)
\left({W\over \rm 10^{50}\,erg}\right) \left({n\over \rm
  1\,cm^{-3}}\right) \left({d\over \rm 1\,kpc}\right)^{-2}.
\label{eq:galactic1}
\end{equation}
Such sources must emerge in an all-sky TeV gamma ray survey performed
with an instrument with the sensitivity of the Milagro experiment\,
\cite{Abdo:2006fq}.

Furthermore, as discussed in the introduction, the position of the knee 
in the cosmic ray spectrum indicates that some sources
accelerate cosmic rays to energies of several PeV. These PeVatrons
therefore produce pionic gamma rays whose spectrum can extend to several
hundred TeV without cutting off. 
For such sources the $\gamma$-ray flux in the TeV energy range can be 
parametrized in terms of a spectral slope
$\alpha_{\gamma}$, an energy $E_{cut,\gamma}$ where the accelerator cuts off and a normalization $k_{\gamma}$
\begin{equation}
\frac{dN_{\gamma}(E_\gamma)}{dE_\gamma}
=k_{\gamma}
 \left(\frac{E_\gamma}{\rm TeV}\right)^{-\alpha_{\gamma}} 
\exp\left(-\sqrt{\frac{E_\gamma}{E_{cut,\gamma}}}\right)
\end{equation}
The estimate in Eq.~(\ref{eq:galactic1}) indicates that fluxes as large as   
$dN_{\gamma}/dE_\gamma \sim 10^{-12}$--$10^{-14}$ 
(TeV$^{-1}$ cm$^{-2}$ s$^{-1}$) can be expected at energies of ${\cal  O}$
(10 TeV). 

To date, the Milagro collaboration has identified 6 such PeVatron candidates. 
Surprisingly, they cluster in star forming regions in the nearby
spiral arms, four in the Cygnus region (MGRO J2019+37, MGRO J2031+41,
MGRO J2043+36 and MGRO J2032+37) and two more (MGRO J1908+06 and
MGRO J1852+01) near galactic longitude of $l=40$\,degrees (where they
are also within the field of view of the Southern telescopes). 

The spectrum of 3 of the sources supports their identification as a
PeVatron. H.E.S.S. observations of MGRO J1908+06 reveal a spectrum
consistent with a $E^{-2}$ dependence from 500\,GeV to 40\,TeV without
evidence for a cutoff\, \cite{hesshotspot}. In a follow-up analysis
the MILAGRO collaboration\, \cite{UHEmilagro} showed that its own data
are consistent with an extension of the H.E.S.S. spectrum to at least
90\,TeV. This is suggestive of pionic gamma rays from a PeVatron whose
cosmic ray beam extends to the knee in the cosmic ray spectrum at PeV
energies. MGRO J2031+41, has been observed \cite{magic2032} by the
MAGIC telescope with a spectrum that is also consistent with
$E^{-2}$. The lower flux measured by MAGIC, which we will
conservatively adopt in our calculations, is likely attributed to the
problem of differentiating the source from the background in a high
density environment like the Cygnus region. Finally, the failure of
Veritas to observe MGRO J2019+37 at lower energies implies that the
slope of the spectrum must be larger than -2.2.

In the end, despite the suggestive evidence, conclusively tracing the
observed gamma rays to pions produced by cosmic-ray accelerators has
so far been elusive. It is one of the main missions of neutrino
telescopes to produce the smoking gun for cosmic-ray production by
detecting neutrinos from charged pions. Neutrino
telescopes detect the Cherenkov radiation from secondary particles
produced by the interactions of high energy neutrinos in highly
transparent and well shielded deep water or ice. They take advantage
of the relatively large cross section of high-energy neutrinos and the
long range of the muons produced. The first kilometer-scale detector,
IceCube, is under construction at the geographic South
Pole\cite{Klein:2008px}. IceCube will consist of 80 kilometer-length
strings, each instrumented with 60 10-inch photomultipliers spaced by
17 m. The deepest module is located at a depth of 2.450\,km so that
the instrument is shielded from the large cosmic ray background at the
surface by approximately 1.5\,km of ice. The strings are arranged at
the apexes of equilateral triangles 125m on a side. The instrumented
detector volume is a cubic kilometer of dark, transparent and sterile Antarctic ice. Each optical sensor consists of a glass
sphere containing the photomultiplier and the electronics board that
digitizes the signals locally using an on-board computer. The
digitized signals are given a global time stamp with residuals
accurate to less than 3\,ns and are subsequently transmitted to the
surface. Processors continuously collect the
time-stamped signals from the optical modules that each function
independently.  The digital messages are sent to a string processor and a
global event trigger. They are subsequently sorted into the Cherenkov
patterns emitted by secondary muon tracks that reveal the direction of
the parent neutrino\cite{Halzen:2006mq}. Operating with 59 out of 80 strings, IceCube's total exposure will reach $1 km^2 year$ in 2009. A more realistic evaluation of its potential to reveal galactic cosmic ray accelerators is therefore timely.

\begin{table}
\begin{tabular}{|l|c|c|c|l|}
\hline
Source & 
$\frac{dN_{\gamma,i}(E_{norm})}{dE_\gamma}$ 
(TeV$^{-1}$ cm$^{-2}$ s$^{-1}$) &
$E_{norm,i}$ (TeV)  & $\alpha_{\gamma,i}$ & $E_{cut,\gamma,i}$ (TeV) \\
\hline
MGRO J2019+37 &  $8.7\times 10^{-15}$ &20 &  2 &  25--800 \\
MGRO J2031+41 & $1.7\times 10^{-14}$  &12 &  2 &  25--800 \\
MGRO J2043+36 & $1.2\times 10^{-14}$  &12 & 1.5 -- 3    & 25--800 \\
MGRO J2032+37 & $0.9\times 10^{-14}$  &12 & 1.5 --3   &  25--800 \\
MGR0 J1908+06 & $8.8\times 10^{-15}$  &20 &  2 &   25--800 \\
MGRO J1852+01 & $5.7\times 10^{-14}$  &12 & 1.5--3  & 25--800 \\
\hline
\end{tabular}
\label{tab:sources}
\caption{Normalization and parameters assumed for the six 
North Hemisphere potential PeVatrons considered in the text.
The value of  $k_{\gamma_i}$ for each source is obtained by 
imposing the normalization condition
given in the second column of the table and are the same as in
Ref.\cite{hkm}.}
\end{table}

We list in table \ref{tab:sources} the spectral parameters for the northern 
hemisphere sources selected as candidate sources.  The choice of sources is
somewhat arbitrary.  We did include the relatively strong source MGRO
J1852+01 despite the fact that it moved at some point below Milagro's
5 sigma pre-trial requirement. On the other hand, we did not include
Geminga and MGRO J2226+60. The latter could actually be supernova
remnant SNR G106.6+2.9; it could also be the Boomerang pulsar wind
nebula.  It is generally assumed that the TeV radiation from pulsar
wind nebulae is electromagnetic in origin so we do not consider them
as potential neutrino sources.
One also has to realize that, after subtracting the sources considered, an excess of TeV gamma rays persists in the Milagro's skymap 
from the general
direction of the Cygnus region\, \cite{Abdo:2008if}. This ``diffuse"
flux most likely originates in unresolved sources that contribute
additional neutrinos.

The neutrino fluxes associated with pionic gamma rays emitted by a source
is directly determined by particle physics; approximately
one $\nu_\mu + \bar\nu_\mu$ pair should accompany every 2 gamma rays.
The exact relation between the gamma ray and neutrino fluxes has been
described in detail in Ref.\cite{Kelner:2006tc}.  Following their 
work we write the  corresponding
neutrino flux at the Earth after oscillations using 
the approximate relations ~\cite{Kelner:2006tc,kappes1}
\begin{equation}
\frac{dN_{\nu_\mu+\bar\nu_\mu,i}(E_\nu)}{dE_\nu} =k_{\nu,i}
\left(\frac{E_\nu}{\rm TeV}\right)^{-\alpha_ {\nu,i}}
\exp\left(-\sqrt{\frac{E_\nu}{E_{cut,\nu,i}}}\right)
\end{equation}
with
\begin{eqnarray}
&& k_\nu=(0.694-0.16 \alpha_\gamma) k_\gamma \nonumber \\ &&
  \alpha_\nu= \alpha_\gamma \nonumber \\ && E_{cut,\nu}=0.59
  E_{cut,\gamma}
\end{eqnarray}
The predicted flux should be robust for the assumption tested that the
sources are molecular clouds illuminated by the beam from nearby
supernova remnants.  Unlike what could be the case for the remnants
themselves, a negligible\cite{gabici} synchrotron component of the TeV
radiation is expected that must be differentiated from the pionic
gamma rays.

Thus the estimates of the neutrino flux are sufficiently quantitative and
the ambiguities associated with the astrophysical parameters suitably
defined, that a quantitative estimate of the response of IceCube is
warranted at this point. Unlike what was done in previous estimates,
we will simulate the detection at the level of the secondary
muons. This allows us to study the influence of energy measurement
which will turn out to be a powerful tool in rejecting the background
of atmospheric neutrinos as we discuss next.

\section{Results}  

The number of events detected by IceCube from a source at 
zenith angle  $\theta_s$ is
\begin{eqnarray}
N_{ev}= t\times N_T \, \int dE_\nu 
dE_\mu^0 dE_\mu^{fin}
&& \Big[
\frac{dN_\nu(E_\nu)}{dE_\nu} 
\times Att_\nu (E_\nu,\theta_s)
\times
\frac{d\sigma_\nu(E_{\nu},E_{\mu}^{0})}{dE_{\mu}^{0}} \nonumber \\
&& 
\!\!\!\!\!\!\!\!\!\!\!\!\!\!\!\!\!\!\!\!\!\!\!\!
+\frac{dN_{\bar\nu}(E_\nu)}{dE_\nu}\times
Att_{\bar\nu} (E_\nu,\theta_s)
\times \frac{d\sigma_{\bar\nu}(E_{\nu},E_{\mu}^{0})}{dE_{\mu}^{0}}
\Big] \nonumber \\
&&
\!\!\!\!\!\!\!\!\!\!\!\!\!\!\!\!\!\!\!\!\!\!\!\!
\!\!\!\!\!\!\!\!\!\!\!\!\!\!\!\!\!\!\!\!\!\!\!\!
\times RR(E_{\mu}^{0},E_{\mu}^{fin}) \times 
A_\mu^{eff}(E_\mu^{fin},\theta_s) 
\label{eq:nevmus}
\end{eqnarray}
The flux sums over equal numbers of neutrinos and antineutrinos of
energy $E_{\nu}$ which are propagated through the Earth; their
zenith-dependent attenuation is described by $Att_\nu
(E_\nu,\theta)$. The neutrinos subsequently interact with a cross
section $\frac{d\sigma_\nu(E_{\nu},E_{\mu}^{0})}{dE_{\mu}^{0}}$ to
yield a secondary muon of energy $E_{\mu}^{0}$. This muon reaches the
detector with an energy $E_{\mu}^{fin}$. The muon propagation is taken
into account by $RR(E_{\mu}^{0},E_{\mu}^{fin})$. The detector
performance is described by its effective area for detecting muons
$A_\mu^{eff}(E_\mu^{fin},\theta_s)$ and the exposure time $t$. $N_T$ is
the target density of the material surrounding the detector. We
emphasize that in order to make a realistic estimate of the
sensitivity of a detector it is essential to describe its performance
at the muon level because it is the muon, not the neutrino energy,
that is measured. This has been overlooked in previous work. In order
to perform the calculations we use the approach introduced in
Ref.\cite{ghm}; the details are collected in the appendix.
  
We subsequently calculate the background of atmospheric neutrinos at
the zenith angle corresponding to the source using the
Honda~\cite{honda} flux which we extrapolate to higher energies to
match the flux from Volkova~\cite{volkova}; the latter is known to
describe the AMANDA data in the energy range of interest here.  At
high energy prompt neutrinos from charm decay contribute.  In order to
estimate the uncertainty associated with the calculation of the charm
flux, we compute the expected number of events for two models of charm
production that bracket predictions in the literature: the
recombination quark parton model (RQPM) developed by Bugaev {\sl et
al}~\cite{rqpm} and the model of Thunman {\sl et al} (TIG)~\cite{tig}
that predicts a smaller rate. We integrate the atmospheric background
over a solid angle $\Omega=\pi/2 (1.6 \delta\psi)^2$ around the
direction of the source.  The angle $\delta\psi$ combines the effects
of the angular resolution of the detector and the size of the
source. Assuming gaussianity, 70\% of the flux of the source is
contained within this angular bin.

We will bin the events in the measured muon energy $E_\mu^{fin}$
assuming a given energy resolution above $E_\mu^{fin}\geq 1$ TeV. Below 1
TeV catastrophic energy loss is small and we assume no energy
resolution; we sum these events in a unique bin.

\begin{figure}
\begin{center}
\includegraphics[height=0.6\textheight]{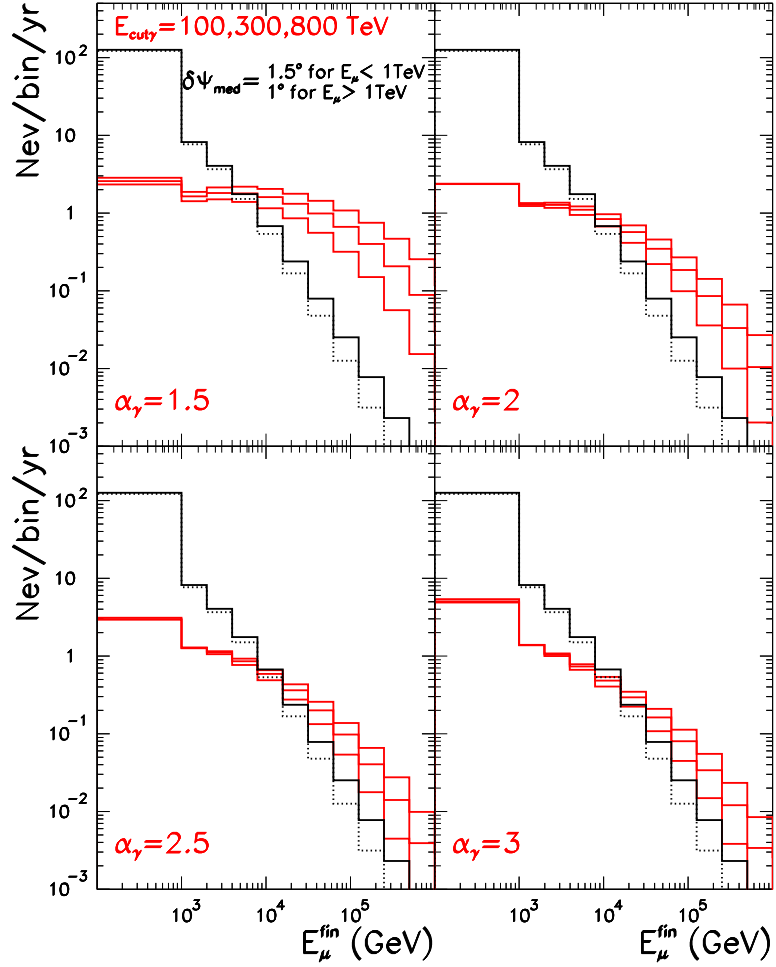}
\end{center}
\caption{Total number of events as a
function of $E^{fin}_\mu$ from the sum of all the sources 
(red histograms) and the corresponding total number of background event 
(steeper black histograms). Background is shown for the two models 
of charm considered: RQPM (full) and TIG (dotted).
In each panel the higher, intermediate and
lower curves for the expected number of events from the sources  
correspond  to $E_{cut,\gamma}=800,300$ and 100 TeV respectively (
assumed to be the same
for all the sources).
In the different panels the $\alpha_\gamma$  for 
MGRO J2043+36, MGRO J2032+37 and MGRO J1852+01 
is given in the figure, while
in all cases we set $\alpha_\gamma=2$ for MGRO J1908+06, MGRO J2031+41 and
MGRO J2019+37. 
In all cases, the background is integrated in an solid angle bin 
$\Omega=\pi/2 (1.6 \delta\psi)^2$  
around the direction of the sources 
with  $\delta\psi=1.5$ degrees for
events with $E_\mu^ {fin}<1$ TeV and  
$\delta\psi=1$ degrees for events with $E_\mu^ {fin}>1$ TeV. }
\label{fig:neventstack}
\end{figure}

Given that the event rates are expected to be a few per candidate
source per year in IceCube, and therefore at the level of the
background of atmospheric neutrinos, the usual procedure is to ``stack" the
sources within the field of view. As an illustration we show in
Fig.\ref{fig:neventstack} the total number of events from the sum of the sources in Table~\ref{tab:sources} and the corresponding
number of background events as a function of $E^{fin}_\mu$. The
results are shown for some representative choices of the angular bin
size and assume an energy resolution of 30\% in $\log(E_\mu^{fin})$
for $E_\mu^{fin}\geq 1$ TeV. The figure illustrates the dependence of
signal rates on variations of the parameters describing the gamma ray
flux were 
we have assumed the same  $E_{cut,\gamma}$ for all the sources.
For the reasons discussed, we
fix $\alpha_\gamma=2$  for MGRO J1908+06, MGRO J2031+41 and
MGRO J2019+37 while for the other three sources, 
MGRO J2043+36, MGRO J2032+37, and MGRO J1852+01, 
we vary  $\alpha_\gamma$ as shown in the figure.
Where the choice of angular bin is concerned, while the
angular resolution of IceCube is better than 1 degree for the
effective area assumed, there is evidence that some sources extend
over a radius of 1.5 degrees, hence our range of bin sizes considered.

\begin{figure}
\begin{center}
\includegraphics[height=0.6\textheight]{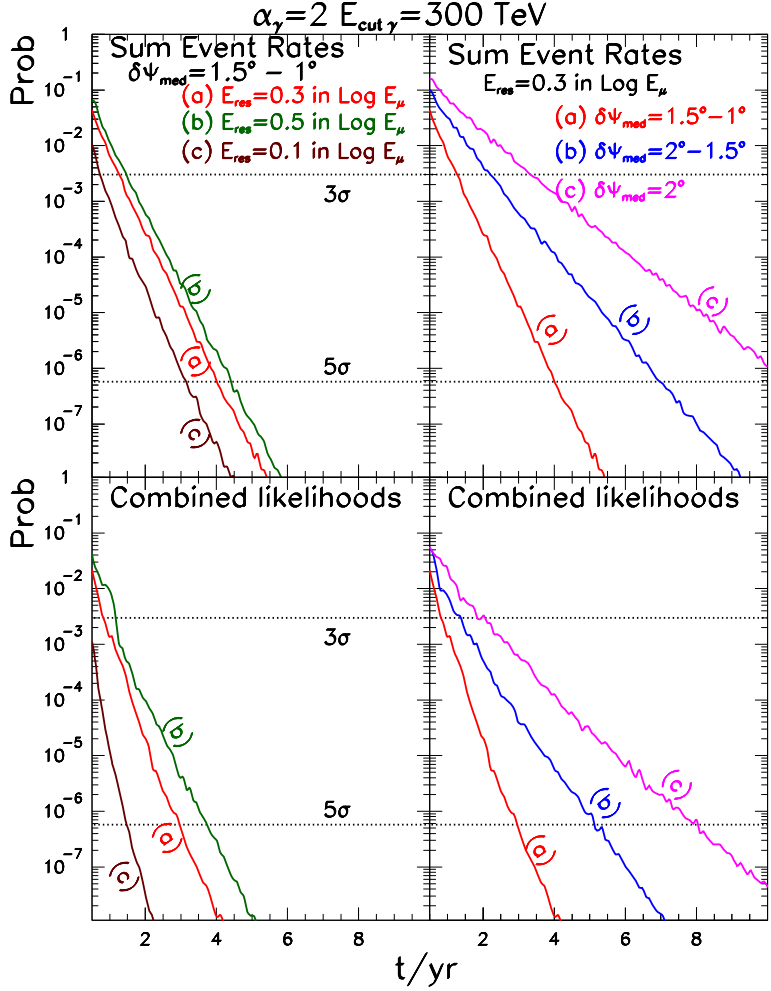}
\end{center}
\caption{Expected significance
for the combined event rate analysis
(upper panels) and for the combined likelihood analysis (lower panels)
as a function of time for different angular bins (right panels)
and energy resolution (left panels) assuming the same 
 $\alpha_\gamma=2$ $E_{cut,\gamma}=300$ TeV
for the 6 sources.
The charm contribution to the atmospheric
background has been computed with the TIG model.
In the left panels the background is integrated over
$\Omega=\pi/2 (1.6 \delta\psi)^2$  with 
$\delta\psi=1.5 (1)$ degrees for
events with $E_\mu^ {fin}<1$ TeV ($E_\mu^ {fin}>1$ TeV). 
The three curves correspond to energy resolution of 
(a) 30\%, (b) 50\% , and (c) 10\% in $\log(E_\mu^{fin})$ for $E_\mu^{fin}> 1$ TeV.
In the right panels an energy resolution of 30\% in $\log(E_\mu^{fin})$ 
and the 
three curves correspond to background integrated over 
a solid angle bin 
$\Omega=\pi/2 (1.6 \delta\psi)^2$  
around the direction of the sources  with 
(a)  
$\delta\psi=1.5$ (1) degrees for
events with $E_\mu^ {fin}<1$ TeV ($E_\mu^ {fin}>1$ TeV),
(b)  $\delta\psi=2$ (1.5) degrees for
events with $E_\mu^ {fin}<1.5$ TeV ($E_\mu^{fin}>1$ TeV),
and (c) $\delta\psi=2$ degrees for all energies. }
\label{fig:probt}
\end{figure}

Given an expected $E_\mu^{fin}$ spectrum for the sources and for the
background we simulate a large number of experiments that are Poisson
distributed around the background, and around signal plus
background. From these sets we obtain the significance or probability
of the observation from the mean likelihood for the observed spectra -- 
assumed to be that of the signal plus background -- to be a random 
fluctuation of
the background. We show in the upper panels of 
Fig.\ref{fig:probt} the expected
significance of the IceCube observation as a function of time for a
range of angular bins and for source parameters $\alpha_\gamma=2$
$E_{cut,\gamma}=300$ TeV. The figure also shows the dependence of the
significance on the energy resolution of the detector.
From the figure we conclude that for these source parameters and 
for the nominal detector resolution 
of 30\% in $\log(E_\mu)$ above 1 TeV and for an angular bin of 
$\delta\psi=1.5$ (1) degrees for events with $E_\mu^ {fin}<1$ TeV 
($E_\mu^ {fin}>1$ TeV), the stacked source analysis yelds a significance of
3$\sigma$ (5$\sigma$) after $\sim$ 1.5 (4.5) years. Widening the
opening of the angular bin by a factor 2 above 1 TeV increases these
times to $\sim$ 3.5 and 11 years. This result is not very much
affected by increasing or decreasing the energy resolution by a factor
of 2. The result does depend on the choice of sources previously
discussed. For instance, if the relatively strong source MGRO J1852+01
were not included, the time for a 5 $\sigma$ observation would
increase by a factor $\sim$ 1.7--2.

\begin{figure}
\begin{center}
\includegraphics[height=0.6\textheight]{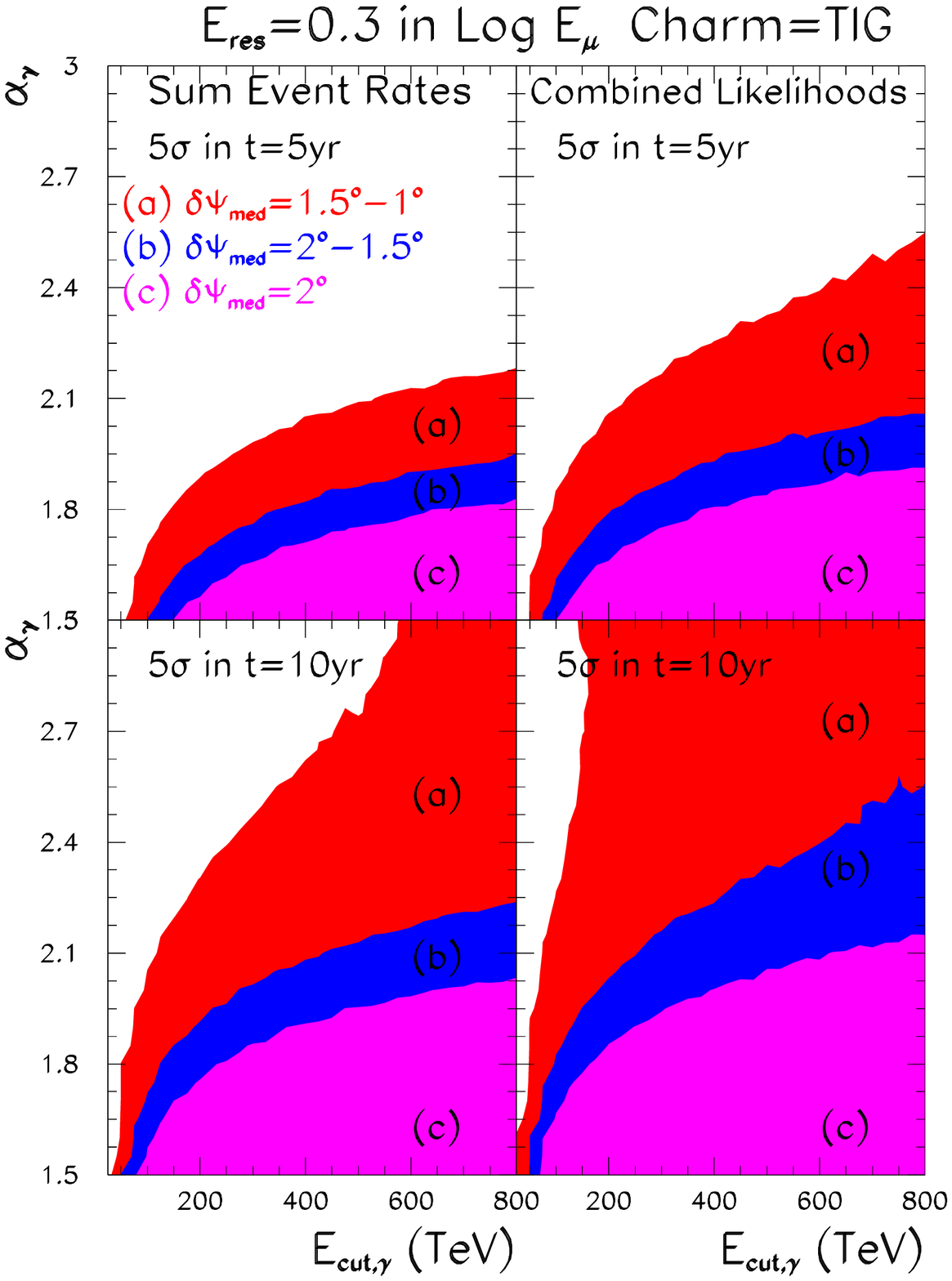}
\end{center}
\caption{Ranges of  $\alpha_\gamma$  and 
$E_{cut,\gamma}$ for which the combined event rate analysis
of the Milagro sources leads to 
a significance of $\geq 5\sigma$ in 5 (upper panels) and 10  (lower panels).
years. In all panels we assume an energy resolution of
30\% in $\log(E^{fin}_\mu)$ and the three full regions (from outer to inner) 
correspond to background integrated over a solid angle bin 
$\Omega=\pi/2 (1.6 \delta\psi)^2$  
around the direction of the sources  with 
(a)  
$\delta\psi=1.5$ (1) degrees for
events with $E_\mu^ {fin}<1$ TeV ($E_\mu^ {fin}>1$ TeV),
(b)  $\delta\psi=2$ (1.5) degrees for
events with $E_\mu^ {fin}<1.5$ TeV ($E_\mu^{fin}>1$ TeV),
and (c) $\delta\psi=2$ degrees for all energies. 
We have assumed 
the same  $E_{cut,\gamma}$ for all the sources while we
fix $\alpha_\gamma=2$  for MGRO J1908+06, MGRO J2031+41 and
MGRO J2019+37.
The left (right) panels corresponds to the analysis of the sum
of the event rates (combined likelihoods) from the sources.}
\label{fig:5sigs}
\end{figure}

Conversely, in Fig.\ref{fig:5sigs}, we show the dependence of the
significance of the observations on the source parameters. The left
panels in the figure show the values of $\alpha_\gamma$ and
$E_{cut,\gamma}$ for which the stacked source analysis leads to a
significance of $\geq 5\sigma$ in 5 or 10 years respectively.  As in
Fig.\ref{fig:neventstack} we have assume the same $E_{cut,\gamma}$ for
all the sources while we fix $\alpha_\gamma=2$ for MGRO J1908+06, MGRO
J2031+41 and MGRO J2019+37.

Alternatively one can simulate experiments around the events expected
from background and signal plus background for each source
individually, and subsequently combine the mean likelihood for the
observed spectra to be a random fluctuation of the
background. Combining the six likelihoods one obtains the probability
and significance of the search.  This method leads to a higher
significance since the expected background is similar for all sources
but the expected number of events is clearly larger for some
sources. In the previous approach the significance of the brighter
sources is diluted. The results following this alternative statistical
approach are shown in the lower panels of Fig.\ref{fig:probt} and in
the left panels of Fig.\ref{fig:5sigs}. For example from  Fig.\ref{fig:probt}
we see that for the nominal detector resolution 
of 30\% in $\log(E_\mu)$ above 1 TeV, and for an angular bin of 
$\delta\psi=1.5$ (1) degrees for events with $E_\mu^ {fin}<1$ TeV 
($E_\mu^ {fin}>1$ TeV), the combined likelyhood analysis can yeld 
a significance of 3$\sigma$ (5$\sigma$) after $\sim$ 1.  (3) years. 

We conclude that within uncertainties in the source
parameters and the detector performance, confirmation that Milagro 
mapped the sources of the Galactic cosmic rays should emerge after 
operating the complete IceCube detector for several years.

\section{A Comment on Observations of Southern Hemisphere Sources}   

\begin{table}
\begin{tabular}{|l|c|c|c|}
\hline
Source & $k_\nu$ (10$^{-12}$ TeV$^{-1}$ cm$^{-2}$ s$^{-1}$ )
& $\alpha_\nu$ & $E_{cut,\nu}$ (TeV) \\
\hline
RX J1713.7-3946 & 15.52 & 1.72 & 1.35 \\
RX J0852.0-4622 & 16.76 &1.78 & 1.19 \\
Vela X & 11.75  & 0.98 & 0.84\\
\hline 
\end{tabular}
\caption{Normalization and parameters assumed for the three
South Hemisphere sources considered in the text.}
\label{tab:shsources}
\end{table}

\begin{figure}
\begin{center}
\includegraphics[height=0.4\textheight]{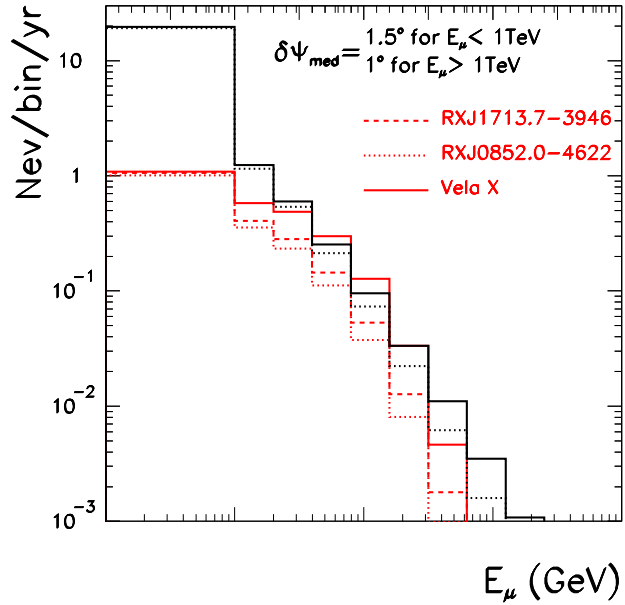}
\end{center}
\caption{Expected event rates from 
RX J1713.7-3946 , RX J0852.0-4622, and Vela X 
 as the atmospheric background  for each source 
(black solid and dotted lines for the two models of charm considered).}
\label{fig:nevents_shsources}
\end{figure}

\begin{figure}
\begin{center}
\includegraphics[width=0.9\textwidth]{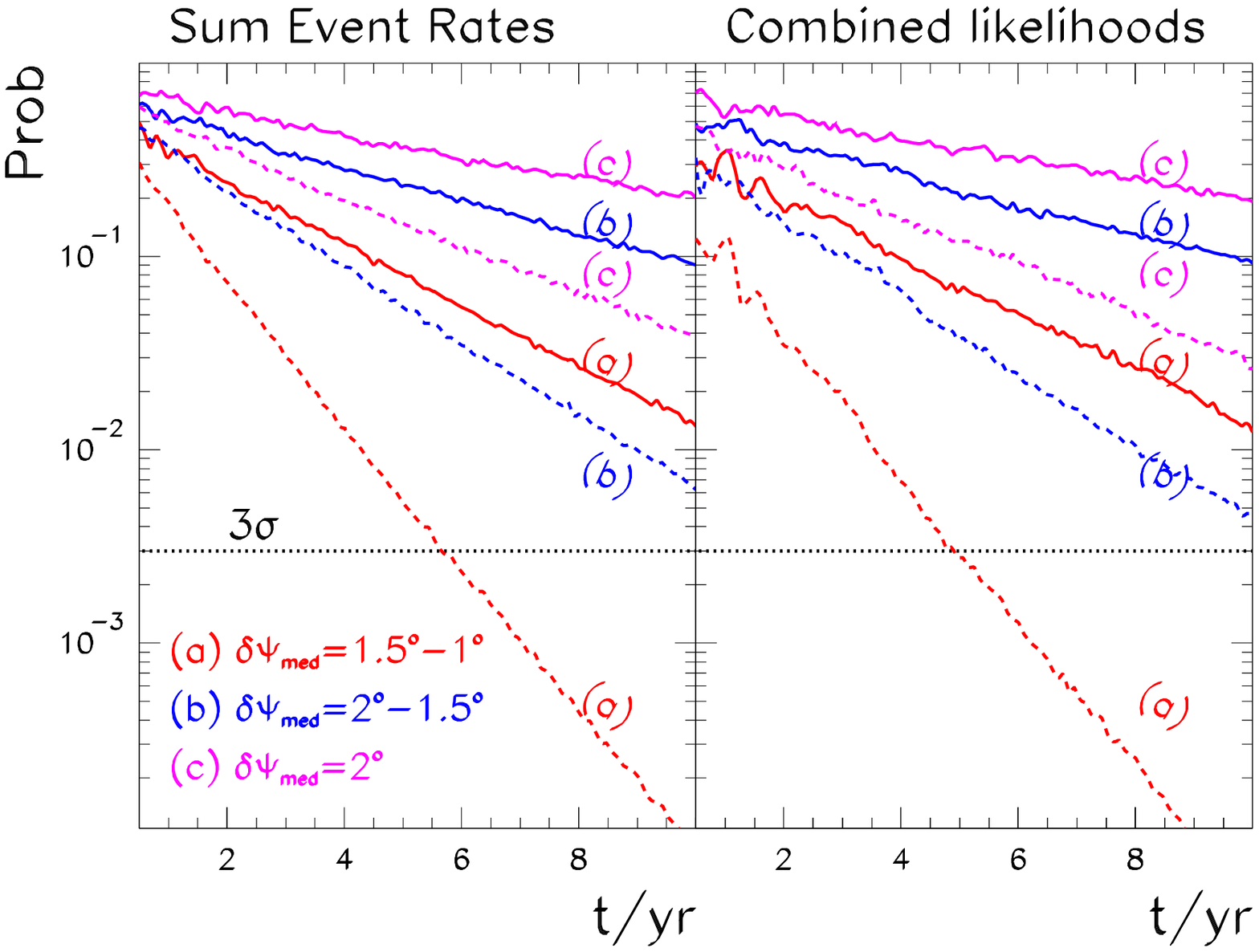}
\end{center}
\caption{Expected significance
from the South Hemisphere sources  
RX J1713.7-3946 and RX J0852.0-4622 (full)  and including also 
Vela X (dashed) as a function of time for different angular bins
from the analysis of the summed event rates (left) and the
combined likelihood functions (right).  
The charm contribution to the atmospheric
background has been computed with the TIG model. 
In both panels we assume an energy resolution of
30\% in $\log(E^{fin}_\mu)$ and the colours 
correspond to background integrated over a solid angle bin 
$\Omega=\pi/2 (1.6 \delta\psi)^2$  
around the direction of the sources  with 
(a)  
$\delta\psi=1.5$ (1) degrees for
events with $E_\mu^ {fin}<1$ TeV ($E_\mu^ {fin}>1$ TeV),
(b)  $\delta\psi=2$ (1.5) degrees for
events with $E_\mu^ {fin}<1.5$ TeV ($E_\mu^{fin}>1$ TeV),
and (c) $\delta\psi=2$ degrees for all energies. }
\label{fig:probtstack_sh}
\end{figure}

For comparison we have studied the expectations for a northern
hemisphere detector, assumed to have a performance identical to
IceCube, viewing supernova remnants in the southern hemisphere. We
took the sources from Ref.\cite{kappes1}; the parameters describing
their neutrino flux are compiled in Table \ref{tab:shsources}. The
event rates predicted are in agreement with those of
Ref.\cite{kappes1} and our results shown in
Fig.\ref{fig:nevents_shsources}. We show in
Fig.\ref{fig:probtstack_sh} the time dependence of the expected
significance for the alternative ways of doing the statistics
previously introduced.

Although the results are disappointing, this is most likely the result
of insufficient information on southern PeVatrons. First, we omitted
the Vela supernova remnant because it has not been detected in TeV
gamma rays. Its pulsar, Vela X, is one of the strongest emitters of
TeV radiation in the sky; the radiation is however believed to be of
electromagnetic origin. Including it in the source list improves the
results as seen in the dashed lines in the figure. It is possible that
some of the unidentified H.E.S.S. sources should be included but in
practice none of them contribute signals at the level of the three
sources considered\cite{kappes1}.  More importantly, before Milagro
none of the sources studied in this paper had been detected in other
wavelengths, or, if observed, had not been pinpointed as candidate
PeVatrons. There have been no TeV all sky observations of the southern
hemisphere. As previously pointed out, IceCube should attempt to
identify such sources in the large background of cosmic ray muons
which it collects at a rate of 1.5\,KHz.

\section{Summary}
We have evaluated as realistically as possible the potential of
IceCube to detect neutrinos in the direction of candidate sources for
accelerators of the Galactic cosmic rays.  In our calculation we make
use of the secondary muon energy determination which allows to
discriminate between the signal and the background from atmospheric
neutrinos. In order to do so we have derived a quantitative
description of the detector performance at the at the level of
secondary muons in the form of a muon effective area.

Our results show that IceCube could identify the sources
in the Milagro sky map as the sources of the Galactic cosmic rays at
the $3\sigma$ level in less than one year and at the $5\sigma$ level
in three years. We have studied the dependence of these expectations on the
ambiguities associated with the  astrophysical 
parameters characterizing the sources and with the detector 
performance. We conclude that 
the absence of evidence after ten years would
reopen the question of the origin of Galactic cosmic rays. 

We briefly comment on the prospects for a northern detector such as
KM3NeT viewing southern sources. They illustrate the potential
importance of an all sky survey of the sourthern hemisphere.

\section*{Acknowledgments}
We thank Dmtri Chirkin and Darren Grant for discussions and  
contributions on the IceCube neutrino effective areas. 
We thank Alexander Kappes and  Aongus O'Murchadha for comments and 
discussions. 
This research was supported by the National Science Foundation
under Grants No.~OPP-0236449 and  PHY-0354776, 
by the U.S.~Department of Energy
under Grant No.~DE-FG02-95ER40896, by the University of
Wisconsin Research Committee with funds granted by the Wisconsin
Alumni Research Foundation and by Spanish Grants
FPA-2007-66665-C02-01, FPA2006-28443-E and  CSD2008-0037.

\section{Appendix: Effective Areas}
Our analysis illustrates the importance of describing the detection of
neutrinos at the level of secondary muons in order to make a reliable
estimate of the response of the detector to predictions.  In this
respect it is important to remark first that the neutrino energy is
not measured by a neutrino telescope. The energy is that of the
secondary muon when triggered by the instrument. The detector's
performance is therefore described by the muon effective area
$A_\mu(E_\mu,\theta)$. The number of observed events from a given
neutrino flux is obtained by convoluting the propagated neutrino flux
with the differential neutrino cross section, the muon propagation and
the muon effective area.

At present the IceCube collaboration exclusively defines the
performance of the detector by the neutrino effective area that simply
yields the number of observed events
after convolution with the neutrino flux. This makes the calculation
of the total event rates simple but does not allow for a realistic
evaluation of the significance of signals which relies on energy
measurement.  Furthermore, the neutrino effective area is not unique
as it depends on the ratio of neutrino and antineutrino fluxes. It is
subject to theoretical ambiguities associated with the neutrino cross
section and muon range.  We remark that all these 
ambiguities are independent of the details of the detector performance.

We here describe how we construct a muon area from the neutrino
area\footnote{The issue has also been recently revisited in
Ref.\cite{ribordy}.}. It is important to point out that the muon area
quantitatively reproduces identical event rates to the neutrino area
at all energies. It is however a theoretical construct only because we
have to make a working identification of the experimentally measured
muon energy and a rather arbitrary variable describing the same
quantity in our calculations.  Following reference~\cite{ghm}, we have
made the identification $E_\mu^{det}= E_\mu^{fin}$.  In doing so we
rely on a procedure that was validated with the last muon area
published by the collaboration~\cite{ghm,ic2004}.

\begin{figure}
\begin{center}
\includegraphics[width=0.6\textwidth]{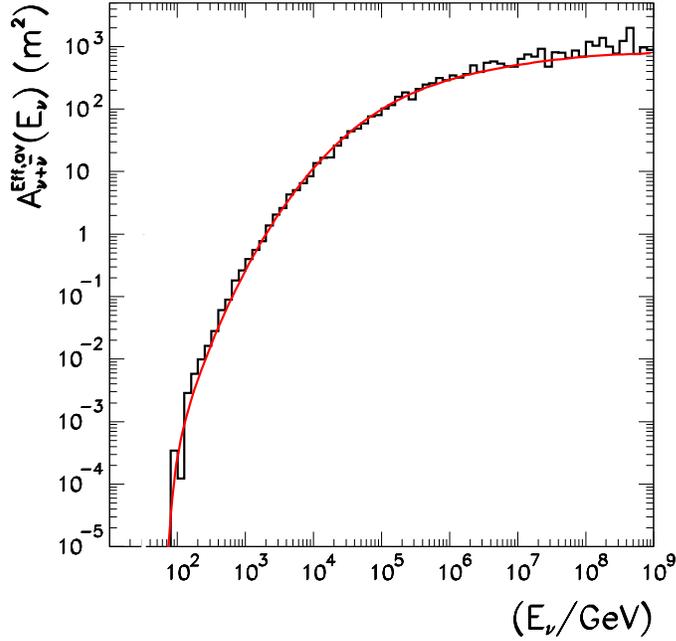}
\end{center}
\caption{The neutrino effective area (averaged over the north hemisphere) 
from the Icecube MCs (black histograms) is compared to the convolution of our
fitted muon effective area (full red line).}
\label{fig:areas}
\end{figure}

We start with the {\sl total} (i.e. angular integrated) neutrino and
antineutrino effective areas of Icecube \cite{ice3},
$A_\nu^{Eff}(E_\nu, bin)$ and $A_{\bar\nu}^{Eff}(E_\nu, bin)$, given
in angular bins labeled $bin\theta$, where $\theta$ is the zenith
angle measured from the vertical. We obtain the corresponding muon
effective area in the same angular bin, $A_\mu^{Eff}(E^{fin}_\mu,
bin\theta)$, from the requirement that both should lead to the same
number of events for an arbitrary flux.  For illustration we plot in
Fig.\ref{fig:areas} the average of neutrino plus antineutrino
effective area averaged also over the northern hemisphere from Icecube
\cite{ice3}
\begin{equation}
A_{\nu+\bar\nu}^{Eff,av}(E_\nu)
=\frac{1}{2\pi}
\sum_{bin} \frac{1}{2} 
\left(A_{\nu}^{Eff}(E_\nu, bin\theta) 
+A_{\bar\nu}^{Eff}(E_\nu, bin\theta)\right)\:.
\label{eq:anuav}
\end{equation}
In terms of the neutrino area the number of events per unit time 
is given  by
\begin{equation}
\frac{N_{ev}}{dt}=\sum_{bin}
\int dE_\nu 
\left(\frac{dN_\nu(E_\nu,bin\theta)}{dE_\nu}
\times A_\nu^{Eff}(E_\nu, bin\theta)
+\frac{dN_{\bar\nu}(E_\nu,bin\theta)}{dE_\nu}
\times A_{\bar\nu}^{Eff}(E_\nu, bin\theta)\right), 
\label{eq:nevnu}
\end{equation}
where 
$dN_{\nu(\bar\nu)}(E_\nu,bin\theta)/dE_\nu$ is the 
{\sl angular-averaged}, energy-differential neutrino (antineutrino) flux 
in the zenith angle bin  $bin\theta$.

Equivalently the number of events can be obtained in terms of the 
muon effective area as:
\begin{eqnarray}
\frac{dN_{ev}}{dt}= N_T\sum_{bin}\, \int dE_\nu 
dE_\mu^0 dE_\mu^{fin}
&& \!\!\!\!\!
\Big[
\frac{d N_\nu(E_\nu,bin\theta)}{dE_\nu} 
\times Att_\nu (E_\nu,bin\theta)
\times
\frac{d\sigma_\nu(E_{\nu},E_{\mu}^{0})}{dE_{\mu}^{0}} \nonumber \\
&& 
\!\!\!\!\!\!\!\!\!\!\!\!\!\!\!\!\!\!\!\!\!\!\!\!
+\frac{d N_{\bar\nu}(E_\nu,bin\theta)}{dE_\nu}\times
Att_{\bar\nu} (E_\nu,bin\theta)
\times \frac{d\sigma_{\bar\nu}(E_{\nu},E_{\mu}^{0})}{dE_{\mu}^{0}}
\Big] \nonumber \\
&&
\times RR(E_{\mu}^{0},E_{\mu}^{fin}) \times 
A_\mu^{eff}(E_\mu^{fin},bin\theta) \:.
\label{eq:nevmu}
\end{eqnarray}
In Eq.~(\ref{eq:nevmu}) $E_\mu^0$ labels the muon energy at its
production point and $E_\mu^{fin}$ its energy when it is detected
after ranging out in the rock and the ice surrounding the detector.
$N_T$ is the number density of targets in the vicinity of the
detector.  $d\sigma_\nu(E_{\nu},E_{\mu}^{0})/ dE_{\mu}^{0}$ is the
differential deep inelastic cross section for which we use the full
expression (without any average inelasticity approximation) obtained
with the CTEQ5 PDFs \cite{CTEQ5}.
$Att_{\nu(\bar\nu)}(E_\nu,bin\theta)$ is a factor which accounts for
the attenuation of the flux due to neutrino (antineutrino) propagation
in the Earth.  It is consistently described by a set of coupled
partial integro-differential cascade equations (see for
example~\cite{reno,ghm} and references therein). The result can be
approximated by~\cite{gaisserbook,ls}
\begin{equation}
Att_{\nu(\bar\nu)}(E_\nu,bin\theta)=
\exp[-X(\theta)(\sigma_{\rm NC}(E)+\sigma_{\rm CC}(E))] \, ,
\label{eq:fluxapp}
\end{equation}
where $X(\theta)$ is the column density of the Earth assuming the
matter density profile of the Preliminary Reference Earth
Model~\cite{PREM}.

$RR(E_{\mu}^{0},E_{\mu}^{fin})$ is the factor which describes the muon
propagation from production to detection:
\begin{equation}
RR(E_{\mu}^{0},E_{\mu}^{fin})
=\int_{0}^{\infty}F
(E_{\mu}^{0},E_{\mu}^{fin},l)dl,
\label{eq:RR}
\end{equation}
where we denote by $F(E^0_\mu,E_\mu^{\rm fin},l)$ the probability that
a muon produced with energy $E_\mu^0$ arrives at the detector with
energy $E_\mu^{\rm fin}$ after traveling a distance $l$.  We compute
the function $F(E^0_\mu,E_\mu^{\rm fin},l)$ by propagating the muons
to the detector taking into account energy losses due to ionization,
bremsstrahlung, $e^+e^-$ pair production and nuclear interactions
according to Ref.~\cite{ls}. In particular we include in
$F(E^0_\mu,E_\mu^{\rm fin},l)$ the possibility of fluctuations around
the average muon energy loss (using the average energy loss would
identify $l$ with the average muon range distance).  Thus in our
calculation we keep $E^0_\mu$, $E_\mu^{\rm fin}$, and, $l$ as
independent variables.  Technically, this is done by numerically
solving the one-dimensional integro-differential equation describing
the muon propagation in matter (see, for example the appendix A of 
Ref.~\cite{ls} and references
therein). 

Comparing  Eq.(\ref{eq:nevnu}) and Eq(\ref{eq:nevmu}) 
we find the relation between the muon and the neutrino areas:
\begin{eqnarray}
A_\nu^{Eff}(E_{\nu},bin\theta)&=&N_T\int
dE^0_\mu dE^{fin}_\mu
Att_\nu (E_\nu,bin\theta)
\times\frac{d\sigma_\nu(E_{\nu},E_{\mu}^{0})}{dE_{\mu}^{0}}
\nonumber \\  
&& 
\times RR(E_{\mu}^{0},E_{\mu}^{fin}) 
\times 
A_\mu^{eff}(E_\mu^{fin},bin\theta),
\label{eq:convol}
\end{eqnarray}
and equivalently for $A_{\bar\nu}^{Eff}(E_{\nu},bin\theta)$.

It is clear from Eq.(\ref{eq:convol}) that without any further
approximation $A_\mu^{eff}(E_\mu^{fin},bin\theta)$ cannot be directly
computed from the neutrino effective areas because they are a double
convolution of the muon areas.  In order to extract the
$A_\mu^{eff}(E_\mu^{fin},bin\theta)$ we assume some parametrization
for its dependence on $E_\mu^{fin}$ and $\theta$ and adjust the
parameters to better describe the neutrino areas.

Overall we find good agreement with a 
very simple functional form:
\begin{equation}
A_\mu^{eff}(E_\mu^{fin},bin\theta)
=2 \pi 
\int_{bin\theta}\; A(E_{\mu}^{fin})\times R(\cos\theta)\;d\cos\theta \; ,
\label{eq:amu1}
\end{equation}
with 
\begin{eqnarray}
&&A(E_{\mu}^{fin}\leq 10^{1.6}\;{\rm GeV})=0   \nonumber \\
&&A(10^{1.6}\;{\rm GeV}\leq E_{\mu}^{fin}\leq 10^{2.8}\;{\rm GeV})
= 0.748 [\log(E_{\mu}^{fin}/{\rm GeV})-1.6]   \nonumber \\
 &&A(10^{2.8}\;{\rm GeV})
\leq E_{\mu}^{fin})
= 0.9 +0.54[\log(E_{\mu}^{fin}/{\rm GeV})-2.8]   \;
\label{eq:amu2}
\end{eqnarray}
and 
\begin{equation}
R(cos\theta)=0.92-0.45\cos\theta
\label{eq:amu3}
\end{equation}
We show  in Fig.\ref{fig:areas} the comparison between the
Icecube effective area and the one which is
obtained from the convolution of the muon effective area in 
Eqs.~(\ref{eq:amu1}--\ref{eq:amu3}).

\end{document}